\documentclass{jpsj2} 
\def\runauthor{J. \textsc{Otsuki}, H. \textsc{Kusunose} and Y. \textsc{Kuramoto}}

\title{
Theory of Crystalline Electric Field and Kondo Effect in Pr Skutterudites}

\author{
Junya \textsc{Otsuki}\thanks{E-mail address: otsuki@cmpt.phys.tohoku.ac.jp},
Hiroaki \textsc{Kusunose}\thanks{E-mail address: kusu@cmpt.phys.tohoku.ac.jp}, 
and Yoshio \textsc{Kuramoto}\thanks{E-mail address: kuramoto@cmpt.phys.tohoku.ac.jp} 
}

\inst{
Department of Physics, Tohoku University, Sendai 980-8578
}

\recdate{\today}

\abst{
Possible Kondo effect in Pr skutterudite is studied with attention to characteristic features of low-lying crystalline electric field (CEF)  levels and the conduction band.
A mechanism for the small CEF splitting between a singlet and a triplet
is proposed as combination of the point-charge interaction
and hybridization of $4f$ with ligand $p$ states.
Provided $4f^3$ configurations dominate over $4f^1$ as intermediate states,
{\it p-f} hybridization favors the triplet,
while point-charge interaction favors the singlet.
For realistic parameters for hybridization as well as $4f^1$ and $4f^3$ levels, 
these singlet and triplet can form a nearly degenerate pseudo-quartet.
It is found that 
one of two spin 1/2 objects composing the pseudo-quartet
has a ferromagnetic exchange, while the other has an antiferromagnetic exchange with conduction electrons. 
The magnitude of each effective exchange depends strongly on a parameter characterizing the triplet wave function under the $T_h$ symmetry.  
It is argued that differences of this parameter among Pr skutterdudites are responsible for the apparent diversity of their physical properties.
Numerical renormalization group is used to derive the renormalization flows going 
toward singlet, doublet, triplet or quaret according to the CEF splitting and exchange interactions.
}

\kword{%
crystalline electric field, hybridization, Kondo effect, filled skutterudite, PrOs$_4$Sb$_{12}$,
PrFe$_4$P$_{12}$
}

\begin{document}
\maketitle

\section{Introduction}

Interactions between conduction electrons and localized degrees of freedom  in metals can lead to profound effects due to multiple particle-hole excitations near the Fermi surface.  
The resistance minimum as a function of temperature was explained first by Kondo\cite{Kondo} who identified the crucial logarithmic term.
In his model,  a magnetic impurity is coupled antiferromagnetically with spins of conduction electrons. 
The essense of physics disclosed by Kondo\cite{Kondo} has far reaching consequences not only in condensed-matter physics, but also in particle physics; 
Kondo effect is the best example of the renormalization effect leading to the confinement and asymptotic freedom\cite{Wilson}.
Since then many variants of Kondo effect have been found.
We mention among others a case 
leading to a non-Fermi liquid ground state \cite{Nozieres}.
The main purpose of this paper is to show that
a new aspect of Kondo-related physics appears
in filled Pr skutterudite compounds where astonishing varieties
of anomalous properties have recently been discovered. 
The properties include
heavy-fermion superconductivity and high-field ordered phase in PrOs$_4$Sb$_{12}$,\cite{Bauer,Aoki} antiferro-quadrupole order in PrFe$_4$P$_{12}$,\cite{PrFeP} metal-insulator and structural phase transitions in PrRu$_4$P$_{12}$.\cite{Sekine,Lee}
While neutron scattering has observed clear CEF transitions in PrOs$_4$Sb$_{12}$,\cite{Maple,Kohgi,kuwahara,gore} only broad quasi-elastic features are visible in PrFe$_4$P$_{12}$ above the temperature of quadrupole order \cite{Iwasa_Fe}.  
On the other hand, intriguing temperature dependence of CEF levels has been observed in PrRu$_4$P$_{12}$ below the metal-insulator transition \cite{Iwasa_Ru}.
For proper understanding of these phenomena, 
interaction effects of 4$f$ electrons with conduction electrons should be taken into account. 
Especially challenging is to identify the parameters that control diversity of phenomena by their slight change.

In this paper, we clarify the importance of 
hybridization between $4f$ and ligand $p$ electrons in giving rise to CEF splittings as well as possible Kondo effect.
It is shown that the observed sequence of CEF levels can be understood as a result of competition between the point-charge mechanism and {\it p-f} hybridization.
Furthermore, it is shown that one of two pseudo-spins forming the lowest four states has antiferromagnetic exchange with the conduction band.
Kondo effect may appear strongly or only weakly depending on 
the wave functions and the level of the triplet CEF states.
We argure that the delicate balance between the CEF splitting and Kondo effect
should be the main reason for the apparently diverse properties of  Pr skutterudites.

\section{Mechanism of CEF splitting in skutterudites}

\subsection{Crystal structure and point-charge model}
In the filled skutterudite structure, each rare-earth ion is surrounded by 8 transition metal ions which form a cube, and 12 pnictogens which form an icosahedron deformed slightly from the regular one.
The Pr site has local symmetry $T_h$ which has no four-fold rotation axis.\cite{Th}
In this symmetry, $4f^2$ Hund's-rule ground states $^3H_4$ split into a $\Gamma_1$ singlet, a $\Gamma_{23}$ non-magnetic doublet, and two $\Gamma_4$ triplets.  Of these, 
$\Gamma_{23}$ corresponds to 
$\Gamma_{3}$ in the point group $O_h$, and two $\Gamma_4$ triplets are written as $\Gamma_4^{(1)}$ and $\Gamma_4^{(2)}$, which are linear combinations of $\Gamma_4$ and $\Gamma_5$ states in $O_h$.
In addition to the Coulomb potential from ligands, hybridization between $4f$ electrons and pnictogen $p$ electrons contributes to CEF splittings.  
Under this crystal symmetry, the CEF potential is written as
\begin{align}
V_{\text{CEF}} 
&	 = A_4 [O_4^0+5O_4^4]+A_6^{\text{c}}[O_6^0-21O_6^4]+A_6^{\text{t}}[O_6^2-O_6^6]
\nonumber \\
& = W\left[ x \frac{O_4}{60} +(1-|x|)\frac{O_6^c}{1260}+y\frac{O_6^t}{30}
 \right],
\label{V_CEF}
\end{align}
in the standard notation\cite{Th}.

In the point charge model, CEF coefficients $A_4$ and $A_6$ in eq.(\ref{V_CEF}) are determined by coordination of the charge and the radial extension of the $4f$ wave function.   For transition ions, the eightfold coordination around Pr gives
\begin{align}	
	 A_4 =\frac{7}{18}\frac{Z_{\rm t}e^2}{d_{\rm t}^5}\langle r^4 \rangle \beta_J , \  
	 A_6^{\text{c}}= -\frac{1}{9}\frac{Z_{\rm t}e^2}{d_{\rm t}^7}\langle r^6 \rangle \gamma_J,
\end{align}	
where $\beta_J$ and $\gamma_J$ are Stevens coefficients, $d_{\rm t}$ is the distance from the origin, and $Z_{\rm t}$ is the effective charge per transition ion.
There is no contribution to
$A_6^{\text{t}}$.   
On the other hand, the deformed icosahedron of pnictogens (X) gives 
\begin{align}	
	 A_4 =
-2f_4(\theta_0)\cdot \frac{7}{16} \frac{Z_{\rm p}e^2}{d_{\rm p}^5}\langle r^4 \rangle \beta_J, \\
	 A_6^{\text{c}}= 
-2f_6^{\text{c}}(\theta_0)\cdot \frac{3}{64} \frac{Z_{\rm p}e^2}{d_{\rm p}^7}\langle r^6 \rangle \gamma_J,\\
	A_6^{\text{t}}=
2f_6^{\text{t}}(\theta_0)\cdot \frac{3}{64} \frac{Z_{\rm p}e^2}{d_{\rm p}^7}\langle r^6 \rangle \gamma_J,
\label{theta}
\end{align}	
where $f_n(\theta_0)$ with $n=4,6$ are numerical factors dependent on the vertical angle $2\theta_0$ of the X-Pr-X triangle as follows:  
\begin{align}
	f_4(\theta_0) &= \frac{1}{2} (\cos 4\theta_0 + \cos^4 \theta_0 + \sin^4 \theta_0), \\
	f_6^{\text{c}} (\theta_0) &= \frac{1}{2} (\cos 4\theta_0 + \cos^4 \theta_0 + \sin^4 \theta_0
	 - 11 \sin^2 \theta_0 \cos^2 \theta_0), \\
	f_6^{\text{t}} (\theta_0) &= \frac{77}{4} (\cos^6 \theta_0 - \sin^6 \theta_0 -\cos 6 \theta_0).
\end{align}
To estimate the CEF potential, 
we use the lattice parameter in PrOs$_4$Sb$_{12}$\cite{Sugawara}, which gives
$d_{\rm t}=4.03\AA, \ d_{\rm p}=3.48\AA$ and $\theta_0=24.6^\circ$.
By interpolating the data of ref.\citen{Freeman},  we use
$\langle r^4 \rangle =3.4a_{\rm B}^4$ and $\langle r^6 \rangle = 19 a_{\rm B}^6$ with $a_{\rm B}$ the Bohr radius.
With trivalent Pr, the effective charges satisfy the relation
$$
3+4Z_{\rm t}+12Z_{\rm p}=0.
$$

\begin{figure}[tb]
\begin{center}
\includegraphics[width=0.9\linewidth,clip]
{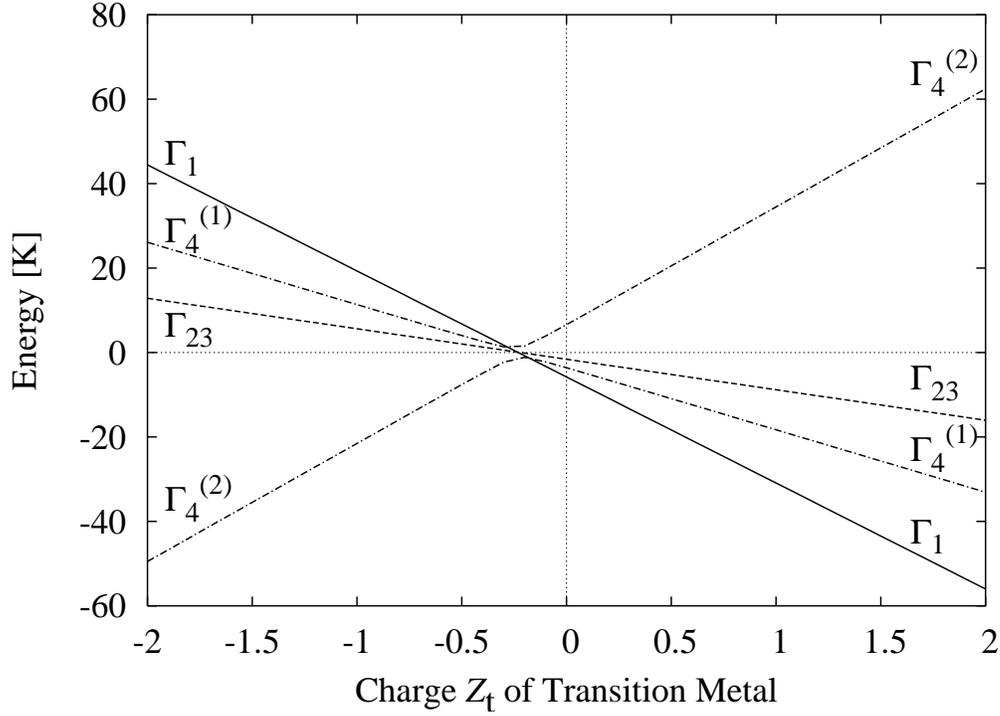}
\end{center}
\caption{CEF level splittings derived from the point-charge model.}
\label{fig:point-charge}
\end{figure}

Figure \ref{fig:point-charge} shows the CEF levels as a function of $Z_{\rm t}$.  
It is seen that with $Z_{\rm t}>0$ the singlet $\Gamma_1$ is favored
as the CEF ground state. 
For slightly negative $Z_{\rm t}$, the CEF potential becomes almost isotropic as pointed out by Takegahara and Harima \cite{Takegahara-Harima}.  The quasi-isotropy is due to high coordination numbers of pnictogens and transition ions, which tend to compensate the anisotropy.

\subsection{CEF splitting by p-f hybridization} 
Another important mechanism for CEF splittings is the covalency effect, or hybridization between localized and ligand orbitals.
Before dealing with Pr skutterudites, let us first consider hybridization between $4f^n$ state and pnictogen molecular orbitals for general $n$. 
We follow the idea of Takahashi and Kasuya \cite{Takahashi-Kasuya}
who studied rare-earth monopnictides.
The large Coulomb repulsion makes it reasonable to neglect the width of the relevant bands in the first approximation. 
Thus the band energy is replaced by its average.
The hybridization is of the form
\begin{align}
	H_{\text{hyb}} &= 
\sum_{\Gamma(\alpha,\beta) \nu \sigma} 
V_{\Gamma{(\alpha,\beta)}}
	 p_{\Gamma{(\alpha)} \nu \sigma}^{\dag} f_{\Gamma{(\beta)} \nu \sigma}
	 +\text{h.c.}, 
\end{align}
where the indices $\alpha,\beta$ 
distinguish different states
with the same irreducible representation $\Gamma$.
A member in the irreducible representation is specified by  $\nu$,
and $\sigma$ denotes a spin component.

We treat hybridization by second order perturbation theory, and assume the following intermediate states:
(i) 4$f^{n-1}$ and an extra electron in vacant states, and
(ii) 4$f^{n+1}$ and extra hole in filled states.
Energy shifts in the case (i)
are given by diagonalization of the following $(2J+1)$-dimensional matrices:
\begin{equation}
	\Delta E_{\Gamma}^{-}(M' ,M)
	= \sum_{\alpha} w_{\alpha} \Delta E_{\Gamma{(\alpha)}}^{-}(M',M),
\end{equation}
where $M, M'$ are $z$-component of $J$, and 
$w_{\alpha}$ denotes the weight of $p$ component in the band with symmetry
$\Gamma{(\alpha)}$, whose energy is represented by $\epsilon_{\Gamma(\alpha)}$.
In the case (ii) we define the matrix
$\Delta E_{\Gamma(\alpha)}^{+}(M' ,M)$ in a similar fashion.
Without multiplet splittings of intermediate states, we obtain
\begin{align}
\Delta E_{\Gamma{(\alpha)}}^{-}(M' ,M) &= -(1-n_{\Gamma(\alpha)})
	 \sum_{\nu \sigma} \sum_{\beta \beta'} 
	 \frac{V_{\Gamma{(\alpha, \beta')}}^{\ast}V_{\Gamma{(\alpha, \beta)}} }
	 {\Delta_{n-1}
	 + \epsilon_{\Gamma(\alpha)}} \nonumber \\
	 & \quad \times 
	 \langle M'| f^{\dag}_{\Gamma{(\beta')} \nu \sigma} 
	 f_{\Gamma{(\beta)} \nu \sigma} | 
	 M \rangle, \label{E-1} \\
\Delta E_{\Gamma{(\alpha)}}^{+}(M' ,M) &= -n_{\Gamma(\alpha)}
	 \sum_{\nu \sigma} \sum_{\beta \beta'} 
	 \frac{V_{\Gamma{(\alpha, \beta)}} V_{\Gamma{(\alpha, \beta')}}^{\ast}}{\Delta_{n+1}
	 + |\epsilon_{\Gamma(\alpha)}|} 
\nonumber \\	 & \quad \times 
\langle 
M'| f_{\Gamma{(\beta)} \nu \sigma} 
	 f_{\Gamma{(\beta')} \nu \sigma}^{\dag} | 
	 M \rangle, \label{E+1}
\end{align}
where 
$\Delta_{n\pm 1}\ (>0)$ are excitation energies to $4f^{n\pm 1}$, and 
$n_{\Gamma(\alpha)}$ is filling of the pnictogen $\Gamma(\alpha)$ state per spin. 
By using the commutation rule of $f$ operators,
we obtain the total energy shift by adding contributions from $4f^{n-1}$ and $4f^{n+1}$ intermediate states:
\begin{align}
&	\Delta E_{\Gamma{(\alpha)}}(M' ,M) 
	= -\sum_{\beta}\frac{|V_{\Gamma{(\alpha, \beta)}}|^2}{\Delta_{n+1} + |\epsilon_{\Gamma(\alpha)}|} 2n_{\Gamma(\alpha)}d_{\Gamma} \ \delta_{M,M'} \nonumber \\ 
& \quad + \left[
	 \frac{ n_{\Gamma(\alpha)}}{\Delta_{n+1} + |\epsilon_{\Gamma(\alpha)}|} - \frac{1-n_{\Gamma(\alpha)}}{\Delta_{n-1} + \epsilon_{\Gamma(\alpha)}} \right]
	  \sum_{\nu \sigma} \sum_{\beta \beta'} V_{\Gamma{(\alpha, \beta)}} V_{\Gamma{(\alpha, \beta')}}^{\ast} \nonumber \\
& \quad \times   \langle 
	M'| f^{\dag}_{\Gamma{(\beta')} \nu \sigma} f_{\Gamma{(\beta)} \nu \sigma} | 
	M \rangle, 
\label{total}
\end{align}
where $d_{\Gamma}$ is the dimension of irreducible representation $\Gamma$. 
The first diagonal term does not contribute to the CEF splitting.

For calculation of matrix elements in eq.(\ref{total}) we need 
a matrix element of the type:
\begin{equation}
	\langle f^{n-1} \gamma' L' M_L' S' M_S' | f_{\Gamma{(\beta)} \nu \sigma}
	 | f^n (LS)JM_J \rangle ,
\end{equation}
where we have rewritten $|M\rangle$ in more detail with $M=M_J$, and
$\gamma'$ is a label to distinguish states with the same $L'$ and $S'$ in $4f^{n-1}$.
The matrix elements between states with plural $4f$ electrons should respect
antisymmetry of the wave function.
This situation is taken care of by a factor
$(f^n \gamma L S \{ | f^{n-1} \gamma' L' S' ,f)$, which is called the coefficient of fractional parentage.\cite{Nielson}  
We obtain \cite{Judd}
\begin{align}
	&\langle f^{n-1} \gamma' L' M_L' S' M_S' | f_{\Gamma{(\beta)} \nu \sigma}
	 | f^n (LS)JM_J \rangle \nonumber \\
	& \quad = (-1)^{n-1}\sqrt{n} (f^n \gamma L S \{ | f^{n-1} \gamma' L' S' ,f) \nonumber \\
	& \quad \times \sum_{m_l M_L M_S} \langle L' M'_L l m_l | L M_L \rangle
	 \langle S' M'_S s \sigma | S M_S \rangle
	 \langle l m_l | l \Gamma{(\beta)} \nu \rangle \langle LM_L SM_S | J M_J \rangle \label{ME-}.
\end{align}
The matrix element in eq.(\ref{total}) can be calculated with use of eq.(\ref{ME-}) 
for any $4f^{n-1}$ configuration. 
It is not necessary to compute the contribution of $4f^{n+1}$ configurations separately.

\begin{figure}[tb]
\begin{center}
\includegraphics[width=0.9\linewidth,clip]
{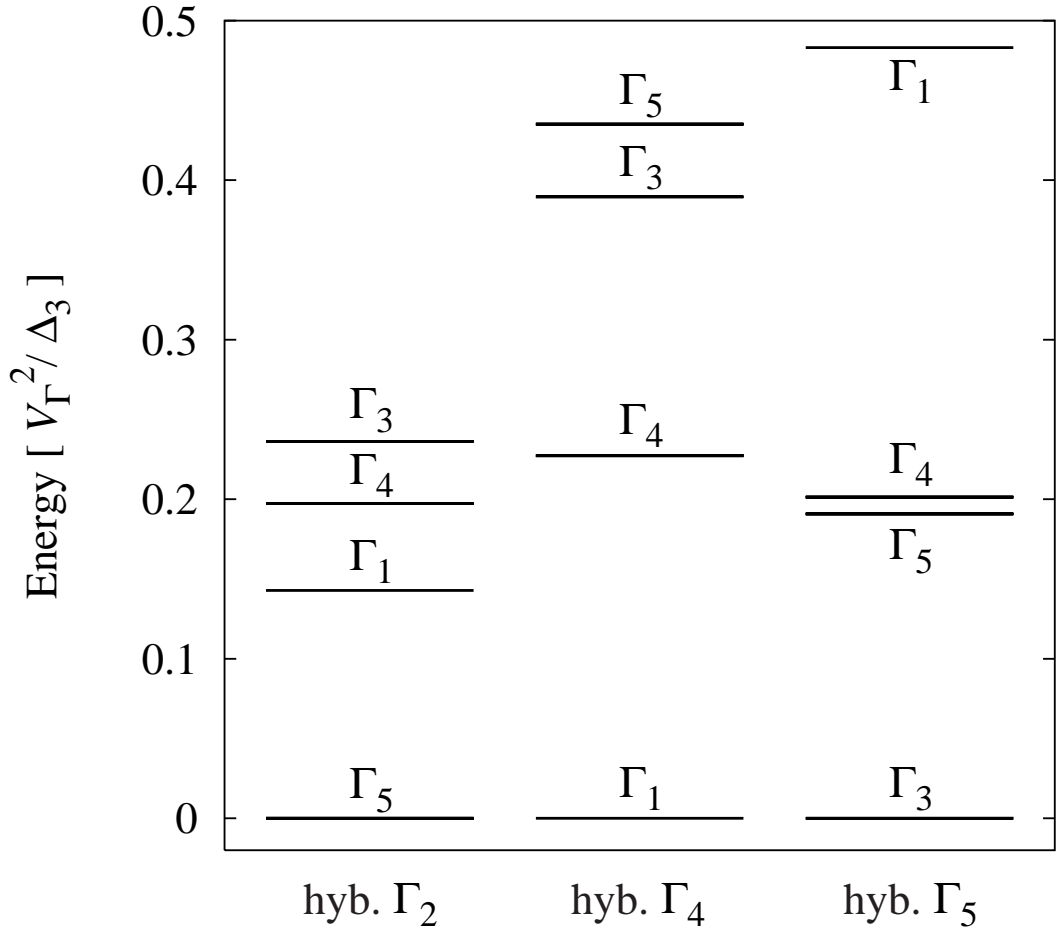}
\end{center}
\caption{CEF level splittings from hybridization with $4f^3$ intermediate states 
in the point group $O_h$.}
\label{fig:energy}
\end{figure}

We  apply this general framework to Pr ion in the point group $O_h$.
Figure \ref{fig:energy} shows CEF level splittings of $4f^2$ states for
different symmetries
$\Gamma_2, \Gamma_4$ and $\Gamma_5$ of hybridization.
We have taken case of $4f^3$ intermediate states, and have assumed
$n_{\Gamma}=1/2$ and 
$\epsilon_{\Gamma}=0$ for each symmetry $\Gamma$.
Since $4f^1$ and $4f^3$ intermediate states give opposite contributions to the level splitting,
the sequence is reversed in the case of $4f^1$ intermediate states.  In reality, the sequence is determined by competition between both intermediate states\cite{Takahashi-Kasuya}.   
In eq.(\ref{total}), the competition appears in the factor with $n_{\Gamma(\alpha)}$.
We have also done calculation where only Hund's-rule ground states in $4f^{2\pm 1}$ are considered as 
intermediate states.
The tendency of opposite contributions from $4f^1$ and $4f^3$ states persist, although the relative distance is not completely the same.  

\subsection{Slater-Koster parameterization}

\begin{figure}[tb]
\begin{center}
\includegraphics[width=0.6\linewidth,clip]{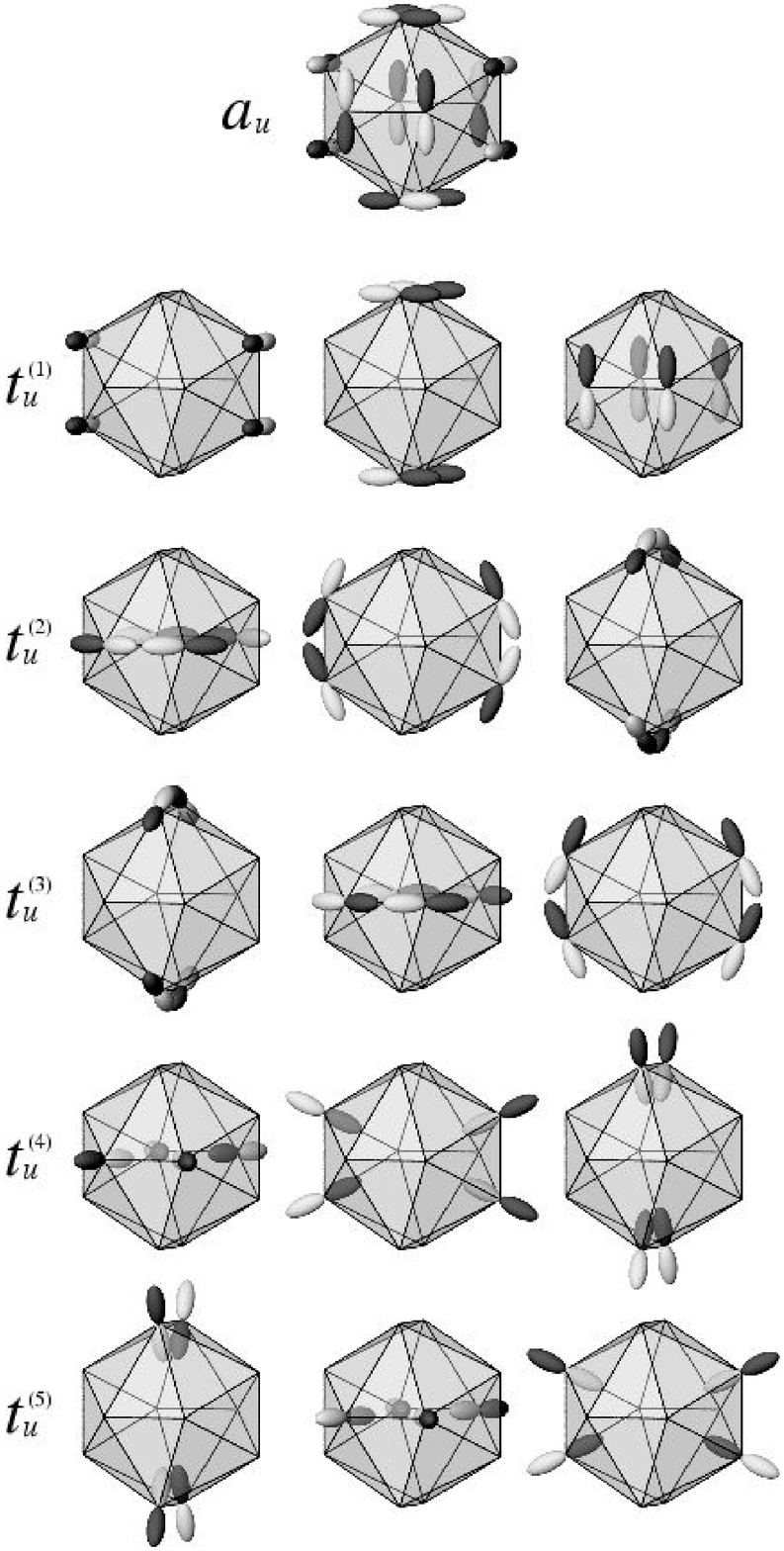}
\end{center}
\caption{Illustration of pnictogen molecular orbitals.  Light and dark parts show the sign of $p$-wave functions at pnictogen sites.}
\label{fig:mo}
\end{figure}

We  proceed to the actual $T_h$ point group around Pr ion in which 
4f electrons hybridize with pnictogen molecular orbitals with the symmetry $a_u \ (=\Gamma_{2})$ and $t_u$.   Hereafter we adopt the Mulliken notation such as $a_u$ for orbital symmetry, and the Bethe notation such as $\Gamma_1$ for a representation with spin-orbit coupling.
Figure \ref{fig:mo} shows relevant molecular orbitals schematically as a combination of atomic $p$ orbitals at pnictogen sites.
The hybridization is parameterized by the Slater-Koster parameters $(pf\sigma)$ and $(pf\pi)$\cite{Takegahara} as shown in 
Table \ref{tab:V}.
It turns out hybridization with $a_u$ comes only from $(pf\pi)$.

\begin{table}[tbh]
	\begin{center}
	\begin{tabular}{|c|cc|}
	 \hline
	  & \multicolumn{2}{c|}{$f:a_u$} \\
	 \hline
	  $p:a_u$ & \multicolumn{2}{c|}{$\sqrt{30}sc(pf\pi)$} \\
	 \hline
	 \hline
	  & $f:t_u^{(1)}$ & $f:t_u^{(2)}$ \\
	 \hline
	  $p:t_u^{(1)}$ & $-\sqrt{3/2} (pf\pi)$ & $\sqrt{5/2} (c^2 -s^2) (pf\pi)$ \\
	  $p:t_u^{(2)}$ & $-\sqrt{3/2} s(5c^2-1)(pf\pi)$ & $\sqrt{5/2}s(2c^2 -s^2)(pf\pi)$ \\
	  $p:t_u^{(3)}$ & $\sqrt{3/2} c(5s^2-1)(pf\pi)$ & $\sqrt{5/2}c(2s^2 -c^2)(pf\pi)$ \\
	  $p:t_u^{(4)}$ & $c(5c^2-3)(pf\sigma)$ & $\sqrt{15}s^2 c(pf\sigma)$ \\
	  $p:t_u^{(5)}$ & $s(5s^2-3)(pf\sigma)$ & $\sqrt{15}s c^2(pf\sigma)$ \\
	 \hline
	 \end{tabular}
	 \end{center}
\caption{Hybridization constant $V_{\Gamma{(\alpha,\beta)}}$ between $f$ and $p$ states with the same symmetry.  The angle $\theta_0$ has been defined below eq.(\ref{theta}), and $s \equiv \sin \theta_0,c \equiv \cos \theta_0$.}  
	 \label{tab:V}
\end{table}

According to band calculation,\cite{Harima} 
the conduction band is formed mainly by $a_u$ orbitals.
On the other hand, two $t_u$ bands are a few eV above the Fermi level and three are a few eV below. 
We neglect contributions from $t_u$ bands to CEF splittings for simplicity.   
We expect the following argument still applies to real Pr skutterudites at least qualitatively by the following two reasons: 
Firstly, the effect of $t_u$ bands is small as compared with the $a_u$ band because of the larger excitation energy. 
Secondly, since $4f^1$ and $4f^3$ intermediate states give opposite contributions to level splittings, the effect of two empty and three filled bands tends to cancel each other.

With only $a_u$ band taken into account,  eq.(\ref{total}) shows that the CEF level sequence is determined by two parameters; occupation of the $a_u$ band and $\Delta_3/\Delta_1$. 
If the latter is less than unity, the sequence becomes the same as that in $4f^3$ intermediate states.
Otherwise the sequence is reversed.
To a good approximation we can assume the half-filled $a_u$ band, and
$\epsilon_{\Gamma}=0$ for the band.
According to available information on PrP \cite{Takahashi-Kasuya},
we obtain the ratio $\Delta_3/\Delta_1=0.6$. 
Hence the level structure due to hybridization alone is dominated by $4f^3$ intermediate states, and should be given as shown in the leftmost part in Fig.\ref{fig:energy}.
However,  we have to consider both point-charge interaction and hybridization for the final sequence of CEF levels.

\subsection{Combination of Coulomb interaction and hybridization}

\begin{figure}[tbh]
\begin{center}
\includegraphics[width=0.9\linewidth,clip]{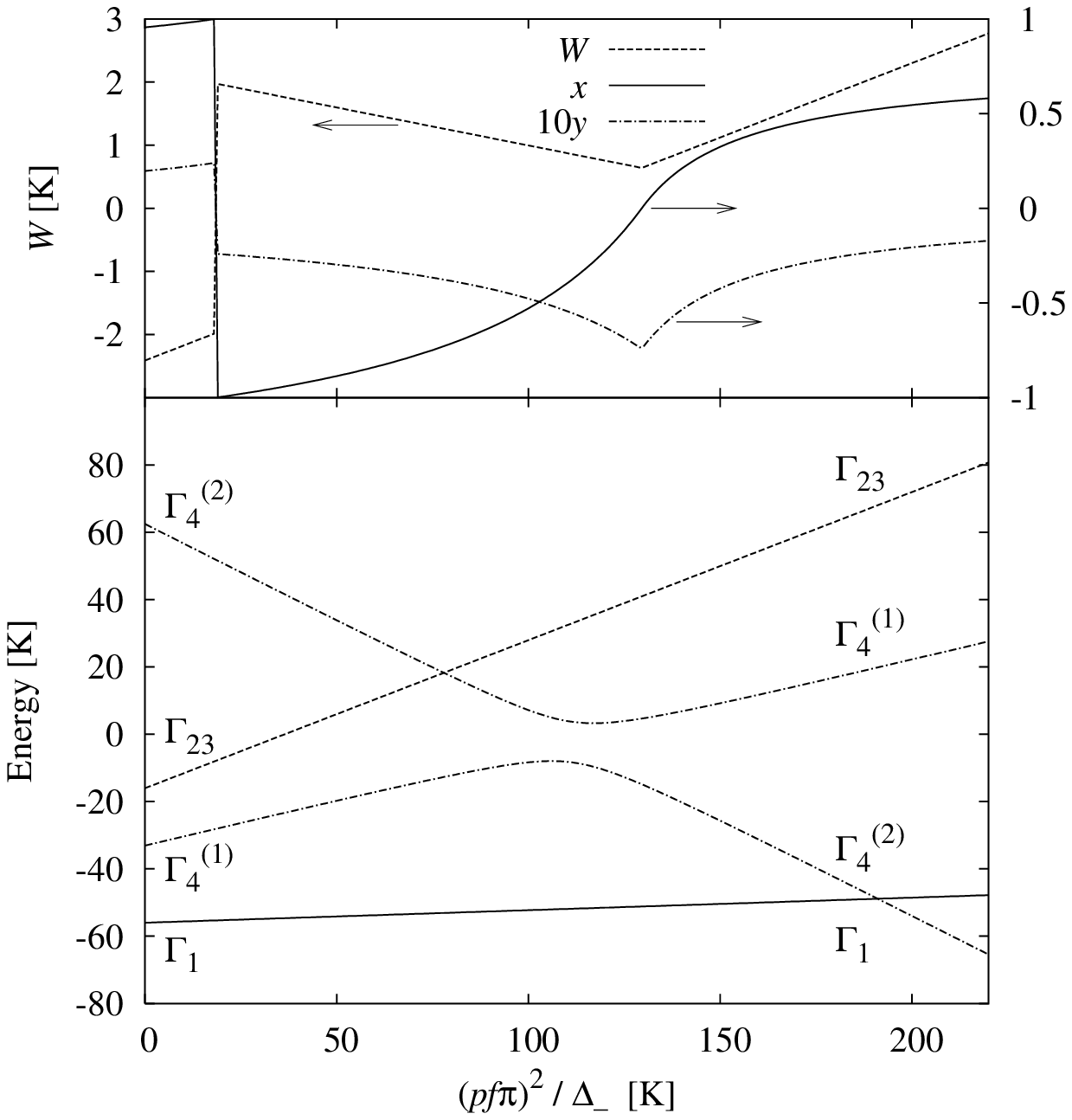}
\end{center}
\caption{CEF parameters $W,x,y$ and
level structures derived from hybridization and point charge potential as a function of 
$(pf\pi)^2/\Delta_-$. Point charges on transition metal and pnictogen are assumed to be +2 and $-11/12$, respectively.}
\label{fig:cef}
\end{figure}

We  combine both contributions to CEF splittings by point-charge
interaction and by hybridization.  
Figure \ref{fig:cef} shows computed results 
as a function of strength of hybridization, $(pf\pi)^2/\Delta_-$ where we have introduced
$$
1/\Delta_-=1/\Delta_3 -1/\Delta_1.
$$
For the point charge parameters we tentatively take $Z_{\rm t}=2$ and $Z_{\rm p}=-11/12$.
We note that in M\"{o}ssbauer experiment on PrFe$_4$P$_{12}$, \cite{Mossbauer}
Fe is reported to be trivalent.  On the other hand, the energy-band calculation suggests a smaller valency for Fe\cite{Harima}.
Overall CEF splittings become larger with larger $Z_{\rm t}$ as seen from Fig.\ref{fig:point-charge}.
As hybridization effect increases, $\Gamma_4^{(2)}$ level is stabilized and crosses the $\Gamma_1$ ground state. 
The parameter $(pf\pi)^2/\Delta_- = 190$K is consistent with the combination of $\Delta_1 \simeq 5$eV, $\Delta_3 \simeq 3$eV and $(pf\pi) \simeq 0.35$eV.   
The $(pf\pi)$ is about 50\% 
larger than that 
obtained for Pr monopnictides.\cite{Takahashi-Kasuya} 
Since the distance $d_{\rm p}$ is smaller than the Pr-P distance of 4.2$\AA$, the larger magnitude is reasonable.
The level repulsion between $\Gamma_4^{(1)}$ and $\Gamma_4^{(2)}$ around $(pf\pi)^2/\Delta_-=110$K is due to mixing of wave functions, and is a characteristic feature in the point group $T_h$.   
In contrast,  a level crossing occurs in the $O_h$ group  
since there is no mixing between $\Gamma_4$ and $\Gamma_5$. 

\section{Exchange interaction in the pseudo-quartet}
We  derive the low-energy effective interaction between $4f$ and conduction states by restricting to the singlet and the lowest triplet for the CEF states.
We shall show that the magnitude of the exchange interaction depends strongly on the nature of triplet wave functions  
even though the strength of hybridization and the CEF splitting are similar.
This difference of effective exchange can be the main reason for the diversity of physical properties in Pr skutterudites.
Although exchange interaction for $f^2$ systems has already been discussed in literature \cite{koga-shiba,cox}, the feature mentioned above has so far escaped theoretical attention. 

\subsection{Second-order exchange interaction}
For simplicity, we neglect the multiplet splittings in intermediate states with $4f^1$ and $4f^3$ configurations.
Only the $a_u$ state is kept for conduction band, which intersects the Fermi surface and is nearly half-filled.
We take the hybridization parameter $V_{2u}$ real and deal with
\begin{equation}
H_{\text{hyb}}
=V_{2u}\sum_{\sigma}f^{\dag}_{\sigma} c_{\sigma} + \text{h.c.},
\end{equation}
where $c_{\sigma}$ annihilates a conduction electron in the Wannier orbital with $a_u$ 
symmetry at the origin,
and $f^{\dag}_{\sigma}$ creates an $4f$ electron with the same ($a_u$) orbital symmetry.
In the second-order perturbation theory, the effective interaction is given by
\begin{equation}
	H_{\text{int}} = P H_{\text{hyb}} \frac{1}{E-H_0} Q H_{\text{hyb}} P,
\end{equation}
where $P$ is the projection operator onto 
$4f^2$ states, and $Q=1-P$.
We have already taken into account such 
part of the second-order hybridization that is diagonal with respect to
the CEF states.  Now we deal with off-diagonal components.

Let us fist consider $4f^1$ as intermediate states, and neglect the multiplet splittings.
Then we obtain 
\begin{align}
	H_{\text{int}} [4f^1]= \frac{V_{2u}^2}{\Delta_1} \sum_{MM'} \sum_{\sigma \sigma'}
	 A(MM' ; \sigma \sigma') \ |M \rangle  \langle M'|\ c^{\dag}_{\sigma} c_{\sigma'},
\label{exch4f1}
\end{align}
where $M$ denotes azimuthal quantum number of $J=4$, and
we have introduced the quantity
\begin{equation}
	A(MM' ; \sigma \sigma') = \sum_{m_l m_{\rm s}} \langle M| f^{\dag}_{\sigma'}
	 | m_l m_{\rm s} \rangle \langle m_l m_{\rm s} | f_{\sigma}|M' \rangle.
\end{equation}
Here $m_l$ and $m_{\rm s}$ denote azimuthal quantum numbers of $l=3$ and $s=1/2$, respectively.

In the case of $4f^3$ intermediate states without multiplet splittings, 
the effective Hamiltonian is given by
\begin{align}
	H_{\text{int}} [4f^3] = -\frac{V_{2u}^2}{\Delta_3} \sum_{MM'} \sum_{\sigma \sigma'}
	 A'(MM' ; \sigma \sigma') \ |M \rangle  \langle M'|\ c^{\dag}_{\sigma} c_{\sigma'},
\end{align}
where we have introduced the quantity
\begin{equation}
	A'(MM' ; \sigma \sigma') = \sum_{f^3} \langle M| f_{\sigma}
	 | f^3 \rangle \langle f^3 | f^{\dag}_{\sigma'} |M' \rangle,
\end{equation}
with summation over all $f^3$ states.
Because of completeness of the intermediate states,
we obtain the relation
\begin{equation}
	A(MM' ; \sigma \sigma')=\delta_{MM'}\delta_{\sigma \sigma'}-A'(MM' ; \sigma \sigma'),
\label{f3tof1}	
\end{equation}
where the first diagonal term is irrelevant. 

Hence consideration of $4f^3$ intermediate states in addition to $4f^1$ 
is accomplished by the replacement
$$
V_{2u}^2/\Delta_1 \rightarrow V_{2u}^2(1/\Delta_1 + 1/\Delta_3)\equiv V_{2u}^2/\Delta,
$$
in eq.(\ref{exch4f1}).
Note that $4f^1$ and $4f^3$ states contribute additively in contrast with the case of CEF splittings. 
Although the bare diagonal terms are already included as CEF splittings, 
higher order potential scattering terms contribute to renormalization of  CEF levels as well as
damping of states.

\subsection{Symmetry analysis in the pseudo-quartet}

Let us first write down the CEF eigenfunctions in $O_h$ in terms of 
$J=4$ states as
\begin{align}
	| \Gamma_1 \rangle = \sqrt{\frac{5}{24}} |+4 \rangle
	 + \sqrt{\frac{7}{12}} |0 \rangle + \sqrt{\frac{5}{24}} |-4 \rangle,
\end{align}
\begin{align}
	| \Gamma_4 ,\pm \rangle &= \mp \sqrt{\frac{1}{8}} |\mp 3\rangle
	 \mp \sqrt{\frac{7}{8}} |\pm 1\rangle, \\
	| \Gamma_4 ,0 \rangle &= \sqrt{\frac{1}{2}}( |+4 \rangle - |-4 \rangle ), \\
	| \Gamma_5 ,\pm \rangle &= \pm \sqrt{\frac{7}{8}} |\pm 3\rangle
	 \mp \sqrt{\frac{1}{8}} |\mp 1\rangle, \\
	| \Gamma_5 ,0 \rangle &= \sqrt{\frac{1}{2}}( |+2 \rangle - |-2 \rangle ).
\end{align}
Following ref.\citen{shiina-aoki}, we represent the lower triplet eigenfunctions in $T_h$ in terms of the functions obtained above:
\begin{align}
	| \Gamma_{\text{t}}, m \rangle = \sqrt{w} | \Gamma_4 ,m \rangle
	 + \sqrt{1-w} | \Gamma_5 ,m \rangle,
\end{align}
where the index $m=0,\pm1$ distinguishes the expectation value of $J_z$,
and the parameter $0<w<1$ measures the weight of $|\Gamma_4 \rangle$.
The other linear combination with a negative coefficient has a higher energy in our CEF potential. 

The direct products of $O_h$ triplets are decomposed as
\begin{align}
	\Gamma_4 \otimes \Gamma_4 = \Gamma_1 + \Gamma_3 + \Gamma_4 + \Gamma_5 
	= \Gamma_5 \otimes \Gamma_5, 
\end{align}
where $\Gamma_4$ has the same symmetry as the magnetic moment. 
Thus we see that both CEF states are magnetic.
On the other hand, the direct products of pseudo-quartets can be decomposed as 
\begin{align}
	(\Gamma_1 \oplus \Gamma_4) \otimes (\Gamma_1 \oplus \Gamma_4)
	 &= \Gamma_1 + 2\Gamma_4 + (\Gamma_1 + \Gamma_3 + \Gamma_4 + \Gamma_5), \\
	(\Gamma_1 \oplus \Gamma_5) \otimes (\Gamma_1 \oplus \Gamma_5)
	 &= \Gamma_1 + 2\Gamma_5 + (\Gamma_1 + \Gamma_3 + \Gamma_4 + \Gamma_5).
\end{align}
Hence, the cross product of 
$\Gamma_1$ and $\Gamma_4$ produces newly two $\Gamma_4$ representations, of which one is time-reversal odd and the other is even.
The odd representation represents the magnetic moment, while the even one the hexadecapole moment.
On the other hand, the pseudo-quartet $\Gamma_1 \oplus \Gamma_5$
 does not produce a new $\Gamma_4$ representation.
Physically, this means that $\Gamma_4$ as the first excited states gives rise to a van Vleck term in the magnetic susceptibility, while $\Gamma_5$ does not.

It is convenient to introduce the effective angular momentum operators in the pseudo-quartet in $T_h$.  In the case of $w=0$ (pure $\Gamma_5$ triplet), 
the magnetic moment within $\Gamma_5$ is the only relevant quantity.  The vector operator in this case is written as  $\mib{X}^{\rm t}$.
On the other hand, another vector operator 
$\mib{X}^{\rm s}$ is necessary to describe the magnetic moment of van Vleck type.  In the matrix representation, we obtain
\begin{align}
	X^{\rm t}_x =\frac{1}{\sqrt{2}} \left( \begin{array}{cccc}
		0 & 0 & 0 & 0 \\
		0 & 0 & 1 & 0 \\
		0 & 1 & 0 & 1 \\
		0 & 0 & 1 & 0
	\end{array} \right) ,\quad
	X^{\rm t}_y =\frac{1}{\sqrt{2}} \left( \begin{array}{cccc}
		0 & 0 & 0 & 0 \\
		0 & 0 & -i & 0 \\
		0 & i & 0 & -i \\
		0 & 0 & i & 0
	\end{array} \right) ,\quad
	X^{\rm t}_z =\left( \begin{array}{cccc}
		0 & 0 & 0 & 0 \\
		0 & 1 & 0 & 0 \\
		0 & 0 & 0 & 0 \\
		0 & 0 & 0 & -1
	\end{array} \right),
\end{align}
\begin{align}
	X^{\rm s}_x =\frac{1}{\sqrt{2}} \left( \begin{array}{cccc}
		0 & -1 & 0 & 1 \\
		-1 & 0 & 0 & 0 \\
		0 & 0 & 0 & 0 \\
		1 & 0 & 0 & 0
	\end{array} \right) ,\quad
	X^{\rm s}_y =\frac{1}{\sqrt{2}} \left( \begin{array}{cccc}
		0 & -i & 0 & -i \\
		i & 0 & 0 & 0 \\
		0 & 0 & 0 & 0 \\
		i & 0 & 0 & 0
	\end{array} \right) ,\quad
	X^{\rm s}_z =\left( \begin{array}{cccc}
		0 & 0 & 1 & 0 \\
		0 & 0 & 0 & 0 \\
		1 & 0 & 0 & 0 \\
		0 & 0 & 0 & 0
	\end{array} \right),
\end{align}
where the basis set is arranged in the order of
$|\Gamma_1 \rangle, |\Gamma_{\rm t} ,+ \rangle, |\Gamma_{\rm t} ,0 \rangle, |\Gamma_{\rm t} ,- \rangle$.
The commutation rules of these operators are given by
\begin{align}
	[ X^{\rm t}_i , X^{\rm t}_j ] = i \epsilon_{ijk} X^{\rm t}_k, \\
	[ X^{\rm s}_i , X^{\rm s}_j ] = i \epsilon_{ijk} X^{\rm t}_k, \\
	[ X^{\rm t}_i , X^{\rm s}_j ] = i \epsilon_{ijk} X^{\rm s}_k, \\
	\text{Tr} X^{\gamma}_i X^{\gamma'}_j = 2\delta_{\gamma \gamma'}\delta_{ij},
\label{commutation-of-X}
\end{align}
where $\gamma,\gamma'$ represent either t or s.

We  project the basis set $|M\rangle$ to the pseudo-quartet.
In terms of two vector operators 
$\mib{X}^{\rm t}$ and $\mib{X}^{\rm s}$, 
we obtain the following form for the effective interaction within the pseudo-quartet
\begin{align}
	H_{\text{int}} &= \frac{V_{2u}^2}{2\Delta} 
	\sum_{\alpha \beta}
	\left( 
	 a_{\rm t} \mib{X}^{\rm t} + a_{\rm s} \mib{X}^{\rm s}
	  \right)
	  \cdot \mib{\sigma}_{\alpha \beta}
	c^{\dag}_{\alpha} c_{\beta}, 
\label{XtXs}	  
\end{align}
where
\begin{align}
	a_{\rm t}=-\frac{10+88w}{1155},\ a_{\rm s}=-\frac{4}{33} \sqrt{\frac{5w}{3}}.
\label{a_i}
\end{align}
In deriving eq.(\ref{a_i}), we used the relation
\begin{align}
	\langle M| f^{\dag}_{\sigma'} | m_l m_{\rm s} \rangle
	= -\sqrt{2} \sum_{m_l' M_L M_S} \langle JM | L M_L S M_S \rangle \nonumber \\
	 \times \langle L M_L | l m_l l m_l' \rangle
	 \langle S M_S | s m_{\rm s} s \sigma' \rangle \langle l m_l' | l \Gamma_2 \rangle,
\end{align}
with 
$\langle l m_l | l \Gamma_2 \rangle = \pm 1/\sqrt{2}$ (for $m_l=\pm2)$.
Both $a_{\rm t}$ and $a_{\rm s}$ are negative, which means ferromagnetic exchange between conduction and $4f$ moments.  The origin of the ferromagnetic sign for $a_{\rm t}$ is traced to the Hund rule involving the spin-orbit interaction; the dominant orbital moment is pointing oppositely from the spin moment.  In other words, the spin exchange is antiferromagnetic as is usual for a hybridization induced exchange.  The sign of $a_{\rm s}$, on the other hand, is not physially meaningful.

\begin{figure}[tb]
	\begin{center}
	\includegraphics[width=0.9\linewidth,clip]{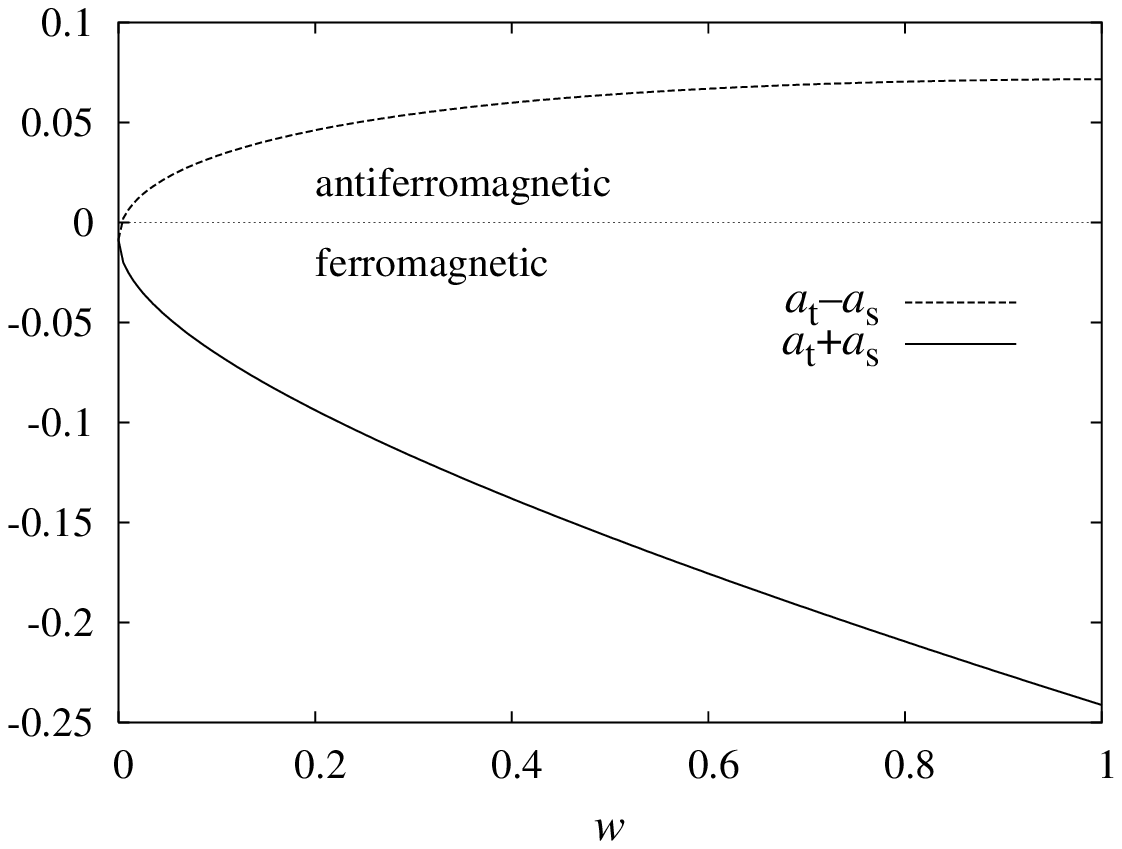}
	\end{center}
	\caption{Coefficients of pseudo-spins in the effective exchange.}
\label{c1c2}	
\end{figure}

The relation to the pseudo-spin representation of ref.\citen{shiina-aoki} is as follows:
$$
\mib{X}^{\rm t} = \mib{S}_1 + \mib{S}_2,\ \mib{X}^{\rm s} = \mib{S}_1 - \mib{S}_2,
$$
where $\mib{S}_1$ and $\mib{S}_2$ are spin 1/2 operators.
It is most natural to interpret the commutation rules  eq.(\ref{commutation-of-X}) in terms of $\mib{S}_1$ and $\mib{S}_2$. 
The exchange interaction in terms of pseudo-spins is given by
\begin{equation}
	H_{\text{int}} = 
	(J_1\mib{S}_1 + J_2\mib{S}_2)\cdot \mib{s}_c,
\label{pseudospin}	
\end{equation}
where 
$
	J_1=(a_{\rm t}+a_{\rm s})V_{2u}^2/\Delta, \ 
	J_2=(a_{\rm t}-a_{\rm s})V_{2u}^2/\Delta
$
and
$$
\mib{s}_c = \frac 12\sum_{\alpha\beta}
c^{\dag}_{\alpha} 
\mib{\sigma}_{\alpha \beta}c_{\beta} .
$$ 
Figure \ref{c1c2} shows the coefficients 
$a_{\rm t}\pm a_{\rm s}$
of pseudo-spins.
It should be noticed 
that $a_{\rm t}-a_{\rm s}$ becomes positive for $w > 0.00324$,  and represents antiferromagnetic exchange $J_2$.
The emergence of antiferromagnetic exchange is due to the particular CEF level structure in Pr skutterudites.  
The antiferromagnetic exchange interaction measured by 
$a_{\rm t}-a_{\rm s}$
is almost negligibe in the pure $\Gamma_5$ case ($w=0$), and becomes an order of magnitude larger as $w$ increases toward unitiy, {\it i.e.}, toward pure $\Gamma_4$.  
It is likely that the Kondo-type behavior seen in PrFe$_4$P$_{12}$ originates from a large value of $w$ together with a small singlet-triplet splitting.

\section{Renormalization group analysis of the singlet-triplet Kondo model}
\label{st-kondo}

In this section we classify all fixed points of the model characterized by the exchange interaction of eq.(\ref{pseudospin}).
Although signs of $J_1$ and $J_2$ are restricted in Pr skutterudites,
we consider here all combinations of signs. 
Possible Kondo effect in systems with a CEF singlet ground state has been discussed from various aspects in literature.\cite{kuramoto,shimizu,yotsuhashi}
Our model relevant to Pr skutterudites has a unique feature of single channel of conduction electrons plus two localized spins with $S=1/2$.
 It will be shown below that the single-triplet CEF system has four
fixed points with different residual entropies.
These fixed points are understood naturally only if we consider renormalization of CEF levels.  Hence we add the CEF terms to  the interaction given by eq.(\ref{XtXs}) and obtain
\begin{equation}
	H_{\text{s-t}} = 
	\epsilon_{\rm t} P_{\rm t}+
	\epsilon_{\rm s} P_{\rm s}+
	\left( 
	 I_{\rm t} \mib{X}^{\rm t} + I_{\rm s} \mib{X}^{\rm s}
	  \right)\cdot \mib{s}_c,
\label{HX}	  
\end{equation}
where
$I_{\rm t} = a_{\rm t}V_{2u}^2/\Delta, \ 
I_{\rm s} = a_{\rm s}V_{2u}^2/\Delta$
and where 
$\epsilon_{\rm t}$ and $\epsilon_{\rm s}$ 
describe the CEF levels. 
Only the difference 
$\Delta_{\rm CEF} \equiv \epsilon_{\rm t} -\epsilon_{\rm s}$ is relevant.
With use of eq.(\ref{pseudospin}), the interaction is alternatively written as
\begin{equation}
	H_{\text{s-t}} =  
\Delta_{\rm CEF} \mib{S}_1\cdot \mib{S}_2 +
	(J_1\mib{S}_1 + J_2\mib{S}_2)\cdot \mib{s}_c.
\end{equation}
We call this model, combined with kinetic energy of conduction electrons, 
the singlet-triplet Kondo model.
The model has interesting features of renormalization because the screening channel is single ($a_u$), while localized spins are two kinds. 

The obvious fixed points of the model are the followings:\\
(i) CEF singlet in the case of $\Delta_{\rm CEF} \gg |J_1|, |J_2|$,\\
(ii) CEF triplet in the case of $-\Delta_{\rm CEF} \gg |J_1|, |J_2|$,\\
(iii) doublet ground state in the case of $-J_1J_2 \gg |\Delta_{\rm CEF}|$,\\
(iv) quartet ground state in the case of both $J_1$ and $J_2$ being renormalized off, and $\Delta_{\rm CEF} =0$.\\
The fixed point (iv) can be regarded as a special case of (i) or (ii) and is accidental. 
The nature of the fixed point (iii), where both the ferromagnetic $J_1$
and antiferromagnetic $J_2$ exchanges are present, 
can be understood as follows:
The component $\mib{S}_2$ will make a singlet by Kondo screening, while the ferromagnetic exchange of other component  $\mib{S}_1$ will be renormalized off.
As a result, the fixed point will be the doublet made up of $\mib{S}_1$. 

\begin{figure}
	\begin{center}
	\includegraphics[width=0.9\linewidth,clip]{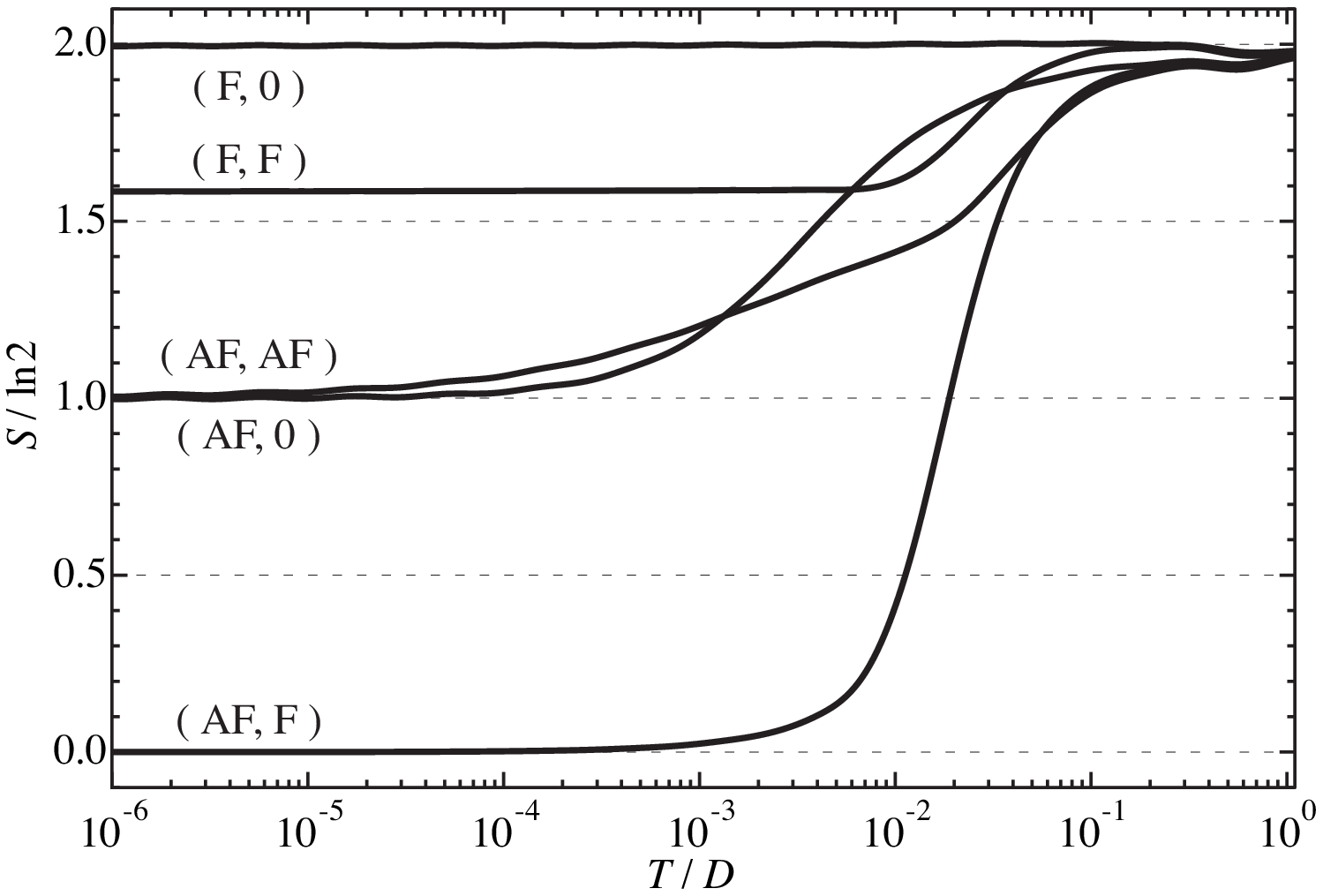}
	\end{center}
	\caption{Entropy of the pseudo-quartet as a function of temperature.
(AF,0) means $J_1>0$ and $J_2=0$, for example.
In the case of (F,F) the residual entropy is $\ln 3$.}
\label{fig:entropy}	
\end{figure}

We have performed explicit calculation using the numerical renormalization group (NRG) for this model.
Figure \ref{fig:entropy} shows the entropy against temperature in the case of  $\Delta_{\rm CEF}=0$ 
for various cases of exchange interactions. 
We used the starting parameters $J_1/D=\pm 0.8, 0$ and $J_2/D=\pm 0.44, 0$ where
$D$ is the half-width of the conduction band with constant density of states.
In Fig.\ref{fig:entropy} 
the entropy in the case of $J_1J_2<0$ (AF,F) goes to zero.
Since we take $\Delta_{\rm CEF}=0$, one might naively expect
the fixed point (iii) in this case with the residual entropy $\ln 2$.
The computed result is naturally interpreted as renormalization of $\Delta_{\rm CEF}$ to a positive value.  
Namely the ground state is connected with the CEF singlet.  

In the case of $J_1<0, J_2 <0$ (F,F), 
the fixed point becomes the triplet (ii), instead of quartet (iv) as might be expected naively.  Again the result is naturally interpreted as renormalization of $\Delta_{\rm CEF}$ which now flows to a negative value.  If one started with the interaction term such as eq.(\ref{pseudospin}) without $\Delta_{\rm CEF}$, the renormalization flow to triplet or singlet would seem mysterious.

There is a tiny critical interaction $J_2 <0$ at which the singlet fixed point is taken over by
the doublet fixed point. 
The critical interaction gives an unstable fixed point at which non-Fermi liquid behavior should be realized.
In contrast with the two-impurity Anderson model,
we do not have a state connecting to two Kondo singlets 
since one of the exchange is ferromagnetic.
Further details of the NRG results will be discussed elsewhere.

In our model of Pr skutterudites with $J_1J_2 <0$ (AF,F),  the renormalization favors the singlet ground state.  Even though the triplet CEF level is lower in second-order perturbation theory,  higher-order terms may reverse the levels.  If the system undergoes the metal-insulator transition, as in PrRu$_4$P$_{12}$,
the singlet may lose the relative stability because the energy gap created in the conduction band will reduce the renormalization effect.  As a consequence, the triplet CEF can be more stabilized as the energy gap increases.   This situation may be relevant to the apparent reversal of CEF levels reported 
in PrRu$_4$P$_{12}$.\cite{Iwasa_Ru}   In the next section, we shall discuss another possibility for the temperature-dependent change of CEF levels. 

\section{Discussion}
\subsection{Comment on other mechanisms for CEF splittings}
In addition to the point-charge mechanism and hybridization, there are other sources for CEF splittings such as anisotropic Coulomb and exchange interactions \cite{Takahashi-Kasuya}.  In the case of rare-earth monopnictides, these other effects involve $5d$ electrons and tend to cancel each other.  In Pr skutterudites,  there is no reason to expect enhancement of anisotropic Coulomb and exchange effects, since the conduction band is composed mainly of $p$ electrons of pnictogens.
Thus we expect that our analysis with two main effects shows realistic origins of CEF splittings in skutterudites.
However,  we have totally neglected intersite interactions of $4f$ electrons which are essential to ordering phenomena.   This aspect should be studied on the basis of CEF level structures obtained here.

\subsection{Interpretation of different behaviors in actual Pr skutterudites}
It is known that the magnetic easy axis is different in PrFe$_4$P$_{12}$ and PrOs$_4$Sb$_{12}$.
Thus the CEF sequence seems also different in both cases.   
In the paramagnetic phase,  PrFe$_4$P$_{12}$ has a sequence of magnetization
$M(100)>M(110)>M(111)$ along each direction of magnetic field \cite{PrFeP}.
The anisotropy in PrFe$_4$P$_{12}$ is consistent with CEF transitions between 
$\Gamma_4$ and $\Gamma_3$ in the notation of $O_h$ group.   
Note that the van Vleck term from $\Gamma_1-\Gamma_4$ transition is isotropic.
Then candidates for the CEF ground state are $\Gamma_1$ and $\Gamma_3$\cite{PrFeP}.
To explain the first-order transition to the quadrupole order, 
Kiss and Fazekas proposed a 
model where $\Gamma_1$ is the ground state\cite{Kiss-Fazekas}.
The sequence of CEF levels appears like the left end of Fig.\ref{fig:cef} where $\Gamma_3$ is written as $\Gamma_{23}$ in the notation of $T_h$. 
In the following discussion we assume the Kiss-Fazekas model for PrFe$_4$P$_{12}$. 

On the other hand,
PrOs$_4$Sb$_{12}$ shows the magnetic anisotropy $M(110)>M(111)>M(100)$ in the paramagnetic phase \cite{Tayama}.  The anisotropy is consistent with CEF transitions between
$\Gamma_5$ and $\Gamma_3$ in the notation of $O_h$ group.   
The sequence near $(pf\pi)^2/\Delta_- \sim 180$K in Fig.\ref{fig:cef} is consistent with this magnetic anisotropy.
In fact,  neutron scattering observed transitions from not only from the $\Gamma_1$ level but from the first excited triplet $\Gamma_4^{(2)}$ \cite{gore}.  The overall intensity suggests that the weight $w$ is small in 
this $\Gamma_4^{(2)}$ states.  

By comparing these facts in PrFe$_4$P$_{12}$ and PrOs$_4$Sb$_{12}$,
we conclude that CEF parameters as shown in Fig.\ref{fig:cef}
vary substantially in Pr skutterudites.
Experimentally, hybridization effect appears stronger in PrFe$_4$P$_{12}$ than in PrOs$_4$Sb$_{12}$.
At first sight, it is paradoxical that CEF levels in PrFe$_4$P$_{12}$ 
look similar to the one in the left end of Fig.\ref{fig:cef}. 
We suggest that the variety of CEF parameters comes from strong cancellation effects between 
hybridization with $4f^3$ and $4f^1$ intermediate states.
If one assumes nearly equal contribution from both intermediate states in PrFe$_4$P$_{12}$, 
we obtain small $|1/\Delta_-| = |1/\Delta_1 - 1/\Delta_3|$.
Thus even though $V_{2u}^2$ itself is large, the CEF levels may be similar to that given by the point-charge model.
As $4f^1$ intermediate states become more dominant than $4f^3$ states, $\Gamma_3$ level becomes stable.
This situation corresponds to negative $\Delta_1$, or reversal of levels in Fig.\ref{fig:energy}.
If such is a case, PrFe$_4$P$_{12}$ may have $\Gamma_3$ as the CEF ground state.

Available information on the ratio $\Delta_3 /\Delta_1$ in actual Pr systems is as follows:
Pr $4f$ spectrum has been observed in X-ray photoemission spectroscopy (XPS) of PrFe$_4$P$_{12}$\cite{Ishii,suga}. 
However, bremsstrahlung isochromat spectroscopy (BIS) for Pr skutterudites has not been carried out to our knowledge.
Data are available for rare-earth metals such
that $\Delta_{n+1}$ in Ce, Pr and Nd are 3.5, 2.1 and 1.7 eV, respectively, and $\Delta_{n-1}$ are $0.3\sim1.9$, 3.3 and 4.7 eV in the same sequence\cite{Lang}.
In the case of monophosphides, it is reported that 
$\Delta_{n+1} /\Delta_{n-1}$ of CeP, PrP and NdP are 2.5, 0.6 and 0.5, respectively \cite{Takahashi-Kasuya}.
The tendency that $\Delta_{n+1} /\Delta_{n-1}$ decreases as $n$ increases is also seen in light rare-earth hexaborides\cite{Mori}. 
Then it seems reasonable to assume $\Delta_3 /\Delta_1 \sim 0.6$ in Pr skutterudites.
Of course, hybridization with $t_u$ ligand orbitals may contribute to the competition to some extent although they were neglected in the present analysis.

Unusual temperature dependence of CEF levels has been found by neutron scatering in PrRu$_4$P$_{12}$ below the metal-insulator transition temperature\cite{Iwasa_Ru}.
In the insulating phase, there appear two inequivalent Pr sites Pr1 and Pr2 with different CEF splittings.   The observation of ref.\citen{Iwasa_Ru} can be qualitatively accounted for if we assume that $(pf\pi)^2/\Delta_-$ increases as temperature decreases below the transition point.
To illustrate the idea, let us suppose in Fig.\ref{fig:cef} that $(pf\pi)^2/\Delta_- \sim 100$K 
above the metal-insulator transition.  Then suppose that $(pf\pi)^2/\Delta_-$  increases by about 10 K in Pr1 sites, and by about 100 K in Pr2 sites.  As a result, Pr2 sites have the triplet CEF ground state, while Pr1 sites keep to the singlet.
However, the triplet changes the character substantially by the small shift of
$(pf\pi)^2/\Delta_-$.
It is an interesting open problem to identify the basic mechanism that causes the change of $(pf\pi)^2/\Delta_-$ in the insulating phase.

\subsection{Dependence of Kondo effect on CEF wave functions}

Even with the same magnitude of $V_{2u}$, the strength of exchange interactions can be much different depending on the CEF wave functions.  As shown in Fig.\ref{c1c2},  the exchange interactions become weak for small $w$,  namely for states close to $\Gamma_5$ states.  This is consistent with the fact that PrOs$_4$Sb$_{12}$ does not show Kondo effect in resistivity \cite{Bauer}.
On the other hand, if the CEF triplet consists of states with $w\sim 1$, i.e., close to $\Gamma_4$ in $O_h$ group, the exchange interactions can be strong.  This is again consistent with the CEF states suggested for PrFe$_4$P$_{12}$, which shows Kondo effect in resistivity\cite{Sato}, and where CEF excitations are hardly visible by neutron scattering in the paramagnetic phase\cite{Iwasa_Fe}.
It should be interesting to derive resisitivity and dynamical susceptibility theoretically for different values of $w$.
The results of such study will be reported in the near future.

\section{Summary}
We have investigated the CEF level structures in Pr skutterudites with
particular attention to the fact that 
hybridization through $4f^{1}$ and $4f^{3}$ intermediate states give opposite contributions to CEF level splittings.
Provided $4f^3$ intermediate states are dominant,
hybridization with the $a_u$ conduction band stabilizes $\Gamma_4^{(2)}$ triplet. 
On the other hand, positive point charges on transition metals 
stabilize $\Gamma_1$ singlet. 
Thus a competition between point charge potential and $p$-$f$ hybridization can lead to the experimentally proposed level structure for PrOs$_4$Sb$_{12}$, i.e., $\Gamma_1$ ground state and $\Gamma_4^{(2)}$ low lying excited states.
The difference between the CEF level structure in PrOs$_4$Sb$_{12}$ and that in PrFe$_4$P$_{12}$ 
is ascribed to difference in competition between $4f^{1}$ and $4f^{3}$ intermediate states rather than difference in strength of hybridization.
It is shown that importance of Kondo effect depends strongly on the character of the CEF triplet next to the singlet.
The predicted tendency according to each CEF wave function
 is consistent with absence of Kondo effect in
PrOs$_4$Sb$_{12}$ in resistivity and strong logarithmic anomaly in PrFe$_4$P$_{12}$. 

\section*{Acknowledgment}
We acknowledge useful discussion with Dr. A. Kiss on relationship between anisotropy of magnetization and CEF levels.

\end{document}